\documentclass[11pt,a4paper]{article}
\pdfoutput=1

\usepackage[style=alphabetic]{biblatex} 
\usepackage{import}
\usepackage{blindtext}

\usepackage{geometry}       
\usepackage{microtype}      
\usepackage{xcolor}         
\usepackage{graphicx}       
\usepackage{amsfonts,amsmath,amssymb,amsthm} 
\usepackage{mathtools}      
\usepackage[en-GB]{datetime2} 

\usepackage{hyperref}   
\hypersetup{
    colorlinks = true,
    linktocpage = true,
    linkcolor = blue,
    citecolor = red,
    urlcolor = blue
}

\AtBeginDocument{
    \numberwithin{equation}{section}
    \numberwithin{figure}{section}
}

\newcommand{\para}{\text{\raisebox{1.5pt}{\fontsize{6pt}{1pt}$\parallel$}}}

\newcommand{\pd}{\partial}
\newcommand{\be}{\begin{equation}}
\newcommand{\ee}{\end{equation}}

\newcommand{\biprod}[2]{
    \langle #1 \, , #2 \rangle
}

\newcommand{\biprodf}[2]{
    \langle #1 \wedge #2 \rangle
}
\newcommand{\triprodf}[3]{
    \langle #1 \wedge [ #2 \wedge #3 ] \rangle
}

\newcommand{\dr}{\mathrm{d}}

\newcommand{\bR}{\mathbb{R}}

\newcommand{\fd}{\mathfrak{d}}
\newcommand{\fg}{\mathfrak{g}}

\newcommand{\CE}{\mathcal{E}}
\newcommand{\CL}{\mathcal{L}}
\newcommand{\CN}{\mathcal{N}}
\newcommand{\CP}{\mathcal{P}}
\newcommand{\CX}{\mathcal{X}}

\begin{document}

\begin{titlepage}

\vfill

\begin{center}
	\baselineskip=16pt  
	
{\Large \bf  \it Unifying Approaches to Chiral Bosons}
	\vskip 1cm
	{\large \bf Alex S.~Arvanitakis$^{a,}$\footnote{\tt alex.s.arvanitakis@vub.be}, Lewis T.~Cole$^{b,}$\footnote{\tt l.t.cole@pm.me}, Ondrej Hulik$^{a,}$\footnote{\tt ondrej.hulik@vub.be}, Alexander~Sevrin$^{a,}$\footnote{\tt alexandre.sevrin@vub.be} and Daniel C.~Thompson$^{a,b,}$\footnote{\tt d.c.thompson@swansea.ac.uk} }
	\vskip .6cm
	{\it  
			$^a$ Theoretische Natuurkunde, Vrije Universiteit Brussel, and the International Solvay Institutes, \\ Pleinlaan 2, B-1050 Brussels, Belgium \\ \ \\
			$^b$ Department of Physics, Swansea University, \\ Swansea SA2 8PP, United Kingdom \\ \ \\}
	\vskip 2cm
\end{center}

\begin{abstract}
\noindent
Chiral bosons, or self-dual $p$-form fields, are ubiquitous in string theoretic contexts but are challenging to treat.  Lagrangian constructions invariably introduce a complexity  be it auxiliary fields or sacrificing Lorentz invariance.  In this note we show how to pass between such different approaches to chiral bosons starting from a Chern Simons point of view to recover formulations of Pasti, Sorokin and Tonin and of Mkrtchyan.  This leads to a novel generalisation of the latter to include non-Abelian chiral bosons in 2-dimensions, and generalisations to include twisted self-duality which are relevant in T-duality symmetric approaches to string theory.    Our approach also shows how global affine symmetries of two- and higher-dimensional chiral bosons emerge from broken Chern-Simons gauge transformations on the boundary. 

\end{abstract}

\vfill

\setcounter{footnote}{0}
\end{titlepage}

\tableofcontents
\newpage

\section{Introduction}

Chiral bosons are an example of self-dual $p$-forms: $p$-form fields $\phi$ in a $(2p + 2)$-dimensional space-time whose $(p+1)$-form field strength $\dr \phi$ is self-dual, i.e.\ $\dr \phi = \star \dr \phi$. Such fields are pervasive within string theory as they are often required to complete multiplets which furnish supersymmetry. Familiar examples include the self-dual $2$-form, living in the $6$-dimensional $\CN = (2, 0)$ multiplet, pertinent to both the M5 brane and the self-dual string. Another example is the self-dual Ramond-Ramond $4$-form of type IIB supergravity. In two dimensions, i.e.\ $p = 0$, the self-duality condition amounts to the chiral boson being purely left-moving, i.e.\ $\pd_- \phi = 0$. These self-dual scalars form a critical component of the ``doubled'' approaches to string theory \cite{Duff:1989tf, Tseytlin:1990nb, Tseytlin:1990va, Hull:2004in, Hull:2006va} in which T-duality is promoted to a manifest symmetry of the worldsheet. Indeed, this behaviour is generic for duality-invariant formalisms and also appears in the context of gauge theory more generally \cite{Henneaux:1988gg, Buratti:2019cbm, Mkrtchyan:2019opf, Bansal:2021bis,Berman:2022dpj}.

Given these motivations, it is highly desirable to have a quantum treatment of chiral bosons. The quantisation of a given theory is made significantly easier if one can find a manifestly Lorentz-invariant Lagrangian, written in terms of Lorentz-covariant objects. However, formulating a Lorentz-invariant Lagrangian for chiral bosons is somewhat challenging, even classically, in essence because the chirality condition is a first order differential equation, whereas one anticipates second order differential equations to arise for bosonic fields. Over the years, many attempts have been made towards finding a suitable Lagrangian for chiral bosons, but most of these formalisms require making some other concession in order to accommodate manifest Lorentz-invariance.

We will provide a more detailed overview of existing approaches to chiral bosons in section \ref{sec:2dCB}, but first, let's summarise the key aspects of some well-known formalisms. Early approaches, such as those of Siegel \cite{Siegel:1983es} and Floreanini and Jackiw \cite{Floreanini:1987as}, sacrificed manifest Lorentz-invariance in order to describe the correct degrees of freedom. Manifest Lorentz-invariance was recovered by Pasti, Sorokin, and Tonin \cite{Pasti:1996vs} at the expense of adding an auxiliary field to the Lagrangian in a non-polynomial manner. More recently, Mkrtchyan \cite{Mkrtchyan:2019opf} added yet another auxiliarly field to the PST action, rendering it polynomial. Unfortunately, the non-polynomial origin of this action is hiding just beneath the surface; in order to demonstrate that the Mkrtchyan action describes a chiral boson, one must leverage a rather mysterious non-polynomial symmetry. As well as improving upon the aforementioned shortcomings, duality-invariant string theory motivates us to include non-abelian fields and twisted self-duality.

Alternative to the two-dimensional approaches, the boundary dynamics of three-dimensional Chern-Simons theory are known to describe chiral bosons \cite{Elitzur:1989nr}. This approach has a few appealing features: the quantisation of Chern-Simons theory has proved very successful, e.g.\ \cite{Elitzur:1989nr, Witten:1988hf, Axelrod:1989xt}; the manifestly Lorentz-invariant action comes equipped with a standard, polynomial symmetry; the non-abelian generalisation is already known; and we will see that the inclusion of a twisted self-duality relation is incredibly simple. While Chern-Simons theory is known to describe chiral bosons, its relationship to the various two-dimensional approaches is less evident. Indeed, one of the key purposes of this note is to clarify these relationships.

In the forthcoming sections, we will demonstrate that, alongside the Floreanini-Jackiw action, both the PST and Mkrtchyan formalisms can be derived from Chern-Simons theory on a manifold with boundary. These derivations will centre around algebraic manipulations of the actions, and (in the abelian case) many of the calculations will also apply to the higher-form versions of the PST and Mkrtchyan actions. Using this relationship, we will also be able to provide the non-abelian and twisted self-duality generalisations of both formalisms. Having done this, we will also have a derivation of the Floreanini-Jackiw action with twisted self-duality, more commonly referred to in the literature as the $\CE$-model where it arises in the context of Poisson-Lie T-duality \cite{Klimcik:1995ux, Klimcik:1996nq, Klimcik:1995dy}. While these generalisations were known for the PST action \cite{Driezen:2016tnz}, we believe that they were previously unknown for the Mkrtchyan action. In both cases, we hope that this derivation will offer a novel perspective on these two-dimensional approaches to chiral bosons. Furthermore, we hope that this note will provide additional motivation to pursue the Chern-Simons approach to chiral bosons.

\section{Approaches to Chiral Bosons in two-dimensions} \label{sec:2dCB}

Let's start by reviewing some of the two-dimensional approaches to chiral bosons which we mentioned in the introduction. Some other approaches to chiral bosons which, for the sake of brevity, we will not explore include those of Henneaux-Teitelboim and Beckaert-Henneaux \cite{Bekaert:1998yp,Henneaux:1988gg}, McClain-Wu-Yu \cite{McClain:1990sx}, Perry-Schwarz \cite{Perry:1996mk}, Sen \cite{Sen:2015nph, Sen:2019qit} and its recent extensions \cite{Barbagallo:2022kbt,Andriolo:2021gen}, and Townsend \cite{Townsend:2019koy, Mezincescu:2022hnb}.

We will consider a bosonic field $\phi$ living on a two-dimensional Lorentzian\footnote{We choose coordinates $(\tau, \sigma)$ and define the metric and orientation by $\dr s^2 = \dr \tau^2 - \dr \sigma^2$ and $\dr^2 \sigma = \dr \tau \wedge \dr \sigma$ respectively. In lightcone coordinates $\sigma^\pm = \frac{1}{2} (\tau \pm \sigma)$, these are given by $\dr s^2 = 4 \, \dr \sigma^+ \dr \sigma^-$ and $\dr^2 \sigma = - 2 \, \dr \sigma^+ \wedge \dr \sigma^-$. The Hodge star acts as $\star \dr \tau = \dr \sigma$, $\star \dr \sigma = \dr \tau$, and $\star \dr \sigma^\pm = \pm \dr \sigma^\pm$. Then, a self-dual field $\dr \phi = \star \dr \phi$ is one for which $\pd_- \phi = 0$. Given a $1$-form $\omega$, we define $\Vert \omega \Vert^2 = g^{\alpha \beta} \omega_\alpha \omega_\beta = \omega_+ \omega_-$.} manifold $\Sigma$. 
Starting with the action for a free non-chiral boson, we may try to incorporate a gauge symmetry $\delta_{\xi} \phi = \xi \pd_- \phi$, such that the only physical content obeys $\pd_- \phi = 0$. This can be done by gauging a chiral conformal symmetry. First, we introduce a gauge field $h$ transforming as\footnote{Weights are naturally $h \equiv h_{++}$ and $\xi \equiv \xi^-$. A complementary perspective is to consider the metric on the worldsheet as $\dr s^2 = 4 \dr \sigma^+ \dr \sigma^- + 4 h (\dr \sigma^+)^2$ with the gauge transformation being the conformal transformations that preserve this metric form.} $\delta_\xi h = \pd_+ \xi + \xi \pd_- h - h \pd_- \xi$, and then we write the action in terms of the would-be-covariant derivatives $\nabla_+ \phi = \pd_+ \phi - h \pd_- \phi$ and $\nabla_- \phi = \pd_- \phi$. Doing this, one arrives at Siegel's gauge invariant action \cite{Siegel:1983es}, 
\begin{equation}
    S_{\text{S}}[\phi] = \int_\Sigma \dr^2 \sigma \, \nabla_+ \phi \nabla_- \phi = \int_\Sigma \dr^2 \sigma \, \big( \pd_+ \phi \pd_- \phi - h (\pd_- \phi)^2 \big) \ . 
\end{equation}
A challenge with this approach is that, even after gauge fixing, the $h$ equation of motion remains as a constraint to be invoked. Whilst the constraint $(\pd_- \phi)^2 \approx 0$ evidently implies $\pd_- \phi \approx 0$, its matrix of first derivatives is  degenerate on the constraint surface making treatment difficult.

Gauge fixing $h = 1$ leads us to the Floreanini-Jackiw [FJ] action \cite{Floreanini:1987as}, 
\begin{equation}
    S_{\text{FJ}}[\phi] = \int_\Sigma \dr^2 \sigma \, \big( \pd_\sigma \phi \pd_- \phi \big) \ , 
\end{equation}
whose equation of motion, although second order, has the general solution 
\begin{equation}
    \pd_- \phi = g(\tau) \ . 
\end{equation}
By virtue of another gauge symmetry\footnote{The Noether charge corresponding to $\tilde{\delta}_{\text{FJ}}$ is zero.}, 
\begin{equation}
    \tilde{\delta}_{\text{FJ}} \phi = h(\tau) \ , 
\end{equation}
the general solution is gauge equivalent to the chirality condition $\pd_- \phi = 0$. An evident downside of this approach, common also to \cite{Henneaux:1988gg, Perry:1996mk}, is that two-dimensional Lorentz invariance is not manifest at the Lagrangian level. Although some one-loop calculations can be done for chiral fields in such a framework, it becomes rather challenging to extend to higher loops.

Notice that this action has another symmetry, 
\begin{equation}
    \hat{\delta}_{\text{FJ}} \phi = \varepsilon (\sigma^+) \ . 
\end{equation}
At first, we might worry that this kills all of the degrees of freedom of our chiral boson. Fortunately, we find that its Noether charge is non-vanishing\footnote{The Noether charge corresponding to $\hat{\delta}_{\text{FJ}}$ is $Q = 2 \int \dr \sigma \, \varepsilon \, \pd_\sigma \phi$.}, and this means that we should not interpret it as a gauge symmetry but rather as a  chiral affine symmetry. This distinction between gauge and affine symmetries will be discussed in greater detail in section \ref{sec:affine}.

Instead of working with the FJ action, one might introduce auxiliary fields so as to restore Lorentz invariance. This is done in the Pasti-Sorokin-Tonin [PST] formalism \cite{Pasti:1996vs} which adopts the action 
\begin{equation}
    S_{\text{PST}}[\phi, f] = \int_\Sigma \dr^2 \sigma \, \bigg( \pd_+ \phi \pd_- \phi - \frac{\pd_+ f}{\pd_- f} (\pd_- \phi)^2 \bigg) \ . 
\end{equation}
Although we have displayed the result with indices explicit, this action can be cast in a manifestly Lorentz invariant fashion, and may be extended to higher form fields. The addition of the field $f$, which is best thought of as the local parameterisation of a closed $1$-form $\omega = \dr f$, is complemented with additional symmetries, 
\begin{align}
    \delta_{\text{PST}} \phi & = \epsilon \frac{\pd_- \phi}{\pd_- f} \ , \quad \delta_{\text{PST}} f = \epsilon \ , \\
    \tilde{\delta}_{\text{PST}} \phi & = \Lambda(f) \ , \quad \tilde{\delta}_{\text{PST}} f = 0 \ . 
\end{align}
Commensurate with this symmetry, the equation of motion for $f$ is automatically satisfied\footnote{Precisely, the $f$ equation of motion is $\frac{\pd_- \phi}{\pd_- f} \pd_- \big( \pd_+ \phi - \frac{\pd_+ f}{\pd_- f} \pd_- \phi \big) = 0$.} given that of $\phi$ which reads 
\begin{equation}
    \pd_- \bigg( \pd_+ \phi - \frac{\pd_+ f}{\pd_- f} \pd_- \phi \bigg) = 0 \ . 
\end{equation}
To see that this encodes a chiral field, it is convenient to introduce a $1$-form $v$ whose components read $v_\pm = \sqrt{\frac{\pd_\pm f}{\pd_\mp f}}$ and a scalar $\chi = v_+ \pd_- \phi$. In terms of these, the equation of motion becomes 
\begin{equation}
    \dr (v \chi) = 0 \ , 
\end{equation}
and the desired self-duality condition now follows as the homogeneous solution $\chi = 0$. Analogous to the general solution of the FJ equation of motion, the inhomogeneous solution $v \chi = \dr \Gamma(f)$ is irrelevant as it can be eliminated by a gauge transformation $\tilde{\delta}_{\text{PST}}(v \chi) = \dr \Lambda(f)$.

Upon gauge fixing the $\delta_{\text{PST}}$-symmetry by setting $f(\tau, \sigma) = \tau$, the PST action reduces exactly to the FJ action, with the residual $\tilde{\delta}_{\text{PST}}$-symmetry becoming $\tilde{\delta}_{\text{FJ}}$. There are some evident downsides, however, to the PST approach: first, the non-polynomial form of the action requires some careful consideration; second, at the functional level, one should restrict to configurations where $\omega = \dr f$ is nowhere vanishing, the existence of which is not a given when taking this approach beyond Minkowski space; third, the PST gauge symmetry appears rather exotic.

Resolving the first of these downsides, Mkrtchyan and collaborators have recently developed an approach which addresses the non-polynomial nature of the PST action \cite{Mkrtchyan:2019opf, Bansal:2021bis, Evnin:2022kqn}. In the spirit of Hubbard-Stratonovich, the PST action can be rendered polynomial by the introduction of an additional scalar field $\alpha$ to give the Mkrtchyan action, 
\begin{equation}\label{eq:RuyLopez}
    S_{\text{M}}[\phi, f, \alpha] = \int_\Sigma \dr^2 \sigma \, \big( \pd_+ \phi \pd_- \phi - 2 \alpha \pd_+ f \pd_- \phi + \alpha^2 \pd_+ f \pd_- f \big) \ . 
\end{equation}
Provided that $\pd_\pm f \neq 0$, one can eliminate $\alpha$ by its equation of motion, $\alpha = \frac{\pd_- \phi}{\pd_- f}$, to recover the PST action. Furthermore, the $\delta_{\text{PST}}$-symmetry is uplifted to 
\begin{equation}
    \delta_{\text{M}} \phi = \epsilon \alpha \ , \quad \delta_{\text{M}} f = \epsilon \ , \quad \delta_{\text{M}} \alpha = \epsilon \frac{\pd_+ \alpha}{\pd_+ f} \ . 
\end{equation}
In \cite{Mkrtchyan:2019opf, Bansal:2021bis, Evnin:2022kqn}, the gauge parameter of this symmetry is redefined according to $\varphi = \epsilon \frac{\pd_+ \alpha}{\pd_+ f}$, such that it is viewed as a shift symmetry on $\alpha$ rather than on $f$. As we can see, while the action is now polynomial, dealing with this symmetry may still prove challenging since it remains non-polynomial. The second symmetry of the PST action also lifts easily to the Mkrtchyan action, 
\begin{equation}
    \tilde{\delta}_{\text{M}} \phi = \Lambda(f) \ , \quad \tilde{\delta}_{\text{M}} f = 0 \ , \quad \tilde{\delta}_{\text{M}} \alpha = \Lambda^\prime (f) \ . 
\end{equation}

It is useful to define the $1$-form $\mu = \dr \phi - \alpha \, \dr f$, whose self-dual component is gauge invariant, 
\begin{equation}
    \delta_{\text{M}} \mu_+ = 0 \ , \quad \delta_{\text{M}} \mu_- = \frac{\epsilon}{\pd_+f} (\pd_+ \mu_- - \pd_- \mu_+) \ , 
\end{equation}
and, in terms of which, the action may be written as 
\begin{equation}
    S_{\text{M}}[\phi, f, \alpha] = \int_\Sigma \big( \mu \wedge \star \mu + 2 \alpha \, \dr f \wedge \dr \phi \big) \ . 
\end{equation}
The existence of this gauge invariant combination $\mu + \star \mu$ is not immediately obvious from the 2d perspective, and indeed it has no analog in the PST formalism. By comparison, we will see later that it appears naturally in the Chern-Simons approach. In \cite{Avetisyan:2022zza}, this gauge invariant combination was essential for extending the free Mkrtchyan formalism presented above to include self-interactions. Indeed, if one wishes to preserve the gauge symmetries of the action, one must be careful to add polynomials of the fields which are independently gauge invariant, and polynomials of $\mu + \star \mu$ are a prime candidate.

Turning to the equations of motion for the Mkrtchyan action (denoting on-shell equivalence by $\simeq$), we have 
\begin{align}
    \frac{\delta S_{\text{M}}}{\delta \alpha} & \simeq 0 \quad \Rightarrow \quad \mu_- \pd_+ f \simeq 0 \ , \\ 
    \frac{\delta S_{\text{M}}}{\delta \phi} & \simeq 0 \quad \Rightarrow \quad \pd_- \mu_+ \simeq 0 \ , \\ 
    \frac{\delta S_{\text{M}}}{\delta f} & \simeq 0 \quad \Rightarrow \quad \mu_- \pd_+ \alpha + \alpha \pd_- \mu_- \simeq 0 \ . 
\end{align}
Assuming that $\pd_\pm f$ are nowhere zero, the $f$ equation of motion is redundant, reflecting the $\delta_{\text{M}}$-symmetry, and the $\phi$ and $\alpha$ equations of motion invoke a flatness, 
\begin{equation}
    \dr \mu = \dr (\alpha \, \dr f) \simeq 0 \ . 
\end{equation}
Solving with $\alpha \, \dr f = \dr \Gamma$, and performing a field redefinition $\phi \to \phi^\prime = \phi - \Gamma$, we then have that $\mu_- = 0$ invokes the desired chirality condition, 
\begin{equation}
    \pd_- \phi^\prime = 0 \ . 
\end{equation}
Shortly, we will encounter the structure we see here, a flatness condition combined with a covariant chirality condition, from another perspective.

\section{The Chern-Simons Approach to Chiral Bosons} \label{sec:CStoCB}

Chiral bosons famously also emerge as the boundary dynamics of Chern-Simons [CS] theory \cite{Elitzur:1989nr, Witten:1996hc, Belov:2006jd}. Consider CS theory on a $3$-manifold $M = \bR \times D$ with the topology of a solid cylinder (the length of the cylinder viewed as the time with coordinate $\tau$, and the disk $D$ parameterised by radial and angular coordinates $\rho$ and $\sigma$). To properly define the action for Abelian CS theory, 
\begin{equation}
    S_{\text{CS}}[A] = \kappa \int_{M} A \wedge \dr A \ , 
\end{equation}
when the manifold $M$ has a boundary, one typically imposes boundary conditions on the connection. These are chosen such that the boundary term in the variation of the action vanishes. In \cite{Elitzur:1989nr}, they chose to impose the boundary condition $A_{\tau} \vert_{\pd M} = 0$, whereas we will deviate and instead impose\footnote{The boundary condition $A_{\tau} \vert_{\pd M} = A_{\sigma} \vert_{\pd M}$ is a self-duality condition $A \vert_{\pd M} = \star (A \vert_{\pd M})$.} $A_{\tau} \vert_{\pd M} = A_{\sigma} \vert_{\pd M}$. Splitting the connection as $A = A_\tau \dr \tau + A^D$, where $A^D$ is a (time-dependent) $1$-form on the disk $D$, the action may be rewritten (after integrating by parts and invoking the boundary condition) as 
\begin{equation}
    S_{\text{CS}}[A] = \kappa \int_{M} \big( 2 A_\tau \dr \tau \wedge \dr^D A^D + A^D \wedge \dr \tau \wedge \pd_\tau A^D \big) + \kappa \int_{\pd M} \! \dr^2 \sigma \, A_\sigma A_\sigma \ . 
\end{equation}
The component $A_\tau$ serves as a Lagrange multiplier enforcing the flatness of $A^D$ which we solve with $A^D = \dr^D \phi = \pd_\rho \phi \, \dr \rho + \pd_\sigma \phi \, \dr \sigma$. Subject to this, the action becomes localised on the boundary after integration by parts, and reads 
\begin{equation}
    S_{\text{CS}} = - \kappa \int_{\pd M} \! \dr^2 \sigma \, \pd_\sigma \phi \pd_- \phi \ , 
\end{equation}
which we recognise as the FJ action. Having demonstrated that Chern-Simons theory indeed describes chiral bosons, we will now consider some augmentations of the action. These will not alter the physical content of the theory, but simply make it more amenable to our future analysis.

Returning to CS theory, we can obtain the self-duality relation as a Neumann type boundary condition by adding a boundary term to the action so that the combined boundary variation takes the form $\int \delta A\wedge(A-\star A)$. (This approach can also be seen in \cite[Appendix A]{Maldacena:2001ss}.) To this end, we define the new functional $S^\prime_\text{CS}$ by 
\begin{equation} \label{eq:Karpov}
    S^\prime_{\text{CS}}[A] = \kappa \int_{M} A \wedge \dr A - \frac{\kappa}{2} \int_{\pd M} \! A \wedge \star A \ , 
\end{equation}
the variation of which is 
\begin{equation}
    \delta S^\prime_{\text{CS}}[A] = \kappa \int_{M} 2 \delta A \wedge \dr A + \kappa \int_{\pd M} \! \delta A \wedge (1 - \star) A \ , 
\end{equation}
and we set the boundary term to zero with $A \vert_{\pd M} = \star (A \vert_{\pd M})$. Of course, in general, CS theory with a boundary is not fully gauge invariant, and instead it is only invariant under those gauge transformations which preserve the boundary condition. Here, including our boundary term, a gauge transformation $\delta_\lambda A = \dr \lambda$ transforms the action as 
\begin{equation}
    \delta_{\lambda} S^\prime_{\text{CS}}[A] = \kappa \int_{\pd M} \! A \wedge (1 - \star) \dr \lambda \ , 
\end{equation}
and thus we only have invariance under self-dual gauge transformations. This may also be understood by considering the gauge transformation of the boundary condition directly.

Morally, this breaking of the gauge symmetry means that would-be pure-gauge modes become propagating edge modes on the boundary. For an alternative perspective, we may restore the gauge invariance by coupling the bulk theory to new boundary degrees of freedom i.e.\ a classical version of anomaly inflow, which has been recently considered in the context of 4d CS theory \cite{Lacroix:2020flf, Benini:2020skc}. Let us define $\chi \in C^{\infty}(\pd M)$ as a boundary field with a gauge transformation $\delta_\lambda \chi = - \lambda$ such that $A^\chi \equiv A + \dr \chi$ is gauge invariant\footnote{It will be helpful for subsequent generalisation to notice that the combination $A^\chi \equiv A + \dr \chi$ can also be thought of as the gauge transformation of the connection where, in the spirit of Stueckelberg, the gauge parameter is promoted to a dynamical field.}.

Evidently, making the replacement $A \to A^\chi$ in our action $S^\prime_{\text{CS}}[A]$ will result in a manifestly gauge invariant action, even with a boundary. Less obvious, however, is that the theory does not depend on the extension of $\chi$ into the bulk, but indeed we find the action 
\begin{equation} \label{eq:Kasparov}
    S_{\text{inv}}[A, \chi] \equiv S^\prime_{\text{CS}}[A^\chi]  = S_{\text{CS}}[A] + \kappa \int_{\pd M} \! \big( A \wedge \dr \chi - \tfrac{1}{2} A^\chi \wedge \star A^\chi \big) \ . 
\end{equation}
The first two contributions here have also been considered in \cite{Arcioni:2002vv} and \cite{Chu:2009ms} to restore the gauge invariance of CS theory with a boundary, and the final term is a stand-alone gauge invariant boundary term. In this larger theory, we have the full gauge freedom with $\dr \lambda$ no longer constrained to be self-dual on the boundary. We can use part of this larger symmetry to fix $\chi = 0$ such that eq.\ \eqref{eq:Kasparov} abbreviates to eq.\ \eqref{eq:Karpov} demonstrating that the physical content of these theories is equivalent.

In this gauge invariant presentation, we recover the chiral boson equations of motion by varying the action, 
\begin{equation}
    \delta S_{\text{inv}}[A, \chi] = \kappa \int_{M} 2 \delta A \wedge \dr A + \kappa \int_{\pd M} \! \delta A \wedge (1 - \star) A^\chi + \kappa \int_{\pd M} \! \delta \chi \big( 2 \dr A - \dr (1 - \star) A^\chi \big) \ . 
\end{equation}
The bulk variation of $A$ (i.e.\ the first term) tells us that $A$ is on-shell flat (which trivially implies that $A^\chi$ is also on-shell flat). Meanwhile, the boundary variation of $A$ gives us the desired self-duality relation on the flat field, $A^\chi \vert_{\pd M} = \star (A^\chi \vert_{\pd M})$. Solving the flatness as $A = \dr (\phi - \chi)$, the boundary equation gives $\dr \phi = \star \dr \phi$. Similar to some of the two-dimensional models, the $\chi$ equation of motion is implied by the other equations of motion, a consequence of the fact that it is gauge trivial.  This ensures that upon gauge fixing $\chi = 0$ one can also disregard its equation of motion and return to the pure CS theory.

At this stage, one might wonder where the self-dual boundary gauge transformations have disappeared to in this presentation. While it might, at first, seem like they have been washed out by the introduction of the edge modes, a simple degrees of freedom counting argument demonstrates that this must not be true. Indeed, the new action $S_{\text{inv}}[A, \chi]$ actually comes with an additional gauge symmetry which acts exclusively on the edge modes. Consider the gauge transformation $\delta_\theta A = 0$ and $\delta_\theta \chi = \theta$. Under this, the action transforms as 
\begin{equation}
    \delta_\theta S_{\text{inv}}[A, \chi] = \int_{\pd M} \! \big( A \wedge (\dr \theta - \star \dr \theta) - \dr \chi \wedge \star \dr \theta \big) \ . 
\end{equation}
So, while the action is not generally invariant under this transformation, if we take $\dr \theta = \star \dr \theta$, then the first term vanishes and the second is a total derivative which we may ignore. These self-dual boundary gauge transformations are precisely the chiral affine symmetries we saw earlier, but this identification will become clearer in the ensuing sections.

The benefit of the edge mode presentation is that this self-dual gauge parameter can be exclusively defined over the boundary, i.e.\ $\theta \in C^\infty (\pd M)$, and acts trivially on the gauge field $A$. Furthermore, the $\lambda$ gauge symmetry is completely unconstrained and has exactly the right degrees of freedom to render $A$ (on-shell) locally trivial everywhere. Conceptually, this makes the transition from a bulk theory to a boundary theory much smoother.

\section{From Chern-Simons to PST and Mkrtchyan}  
Having seen how to obtain the FJ action through manipulations of CS theory, our next goal is to understand how the PST and Mkrtchyan actions can also be recovered. The idea here is \textit{not} to single out the $A_\tau$ component as auxiliary, but instead to introduce a more general decomposition of the bulk gauge field.

The arguments we present here are, to a large extent, independent of dimension, and can equally be applied to self-dual $2$-forms in 6d as to chiral scalars in 2d. Accordingly, we will leave the dimension fairly general, and work in $(2n + 1)$-dimensional Chern-Simons theory for an Abelian $n$-form field $A \in \Omega^n (M)$ on a manifold with a boundary $\pd M \equiv \Sigma$. To restore complete gauge invariance under $\delta_\lambda A = \dr \lambda$ for $\lambda \in \Omega^{n-1} (M)$, we again invoke a Stueckelberg compensator field $\chi \in \Omega^{n-1} (\pd M)$ which transforms with a shift symmetry $\delta_\lambda \chi = - \lambda$. We then use the action $S_{\text{inv}}[A, \chi]$ \eqref{eq:Kasparov}, now understood in this general setting.

One further comment on the number of dimensions must be made. When $\Sigma$ is four-dimensional with Euclidean signature, the bulk Chern-Simons term $A \wedge \dr A$ for a $2$-form $A \in \Omega^2 (M)$ is a total derivative and eq.\ \eqref{eq:Kasparov} immediately simplifies to a boundary action, 
\begin{equation}
    \frac{\kappa}{2} \int_{\pd M} \! \big( A \wedge A + 2 A \wedge \dr \chi - A^\chi \wedge \star A^\chi \big) \ . 
\end{equation}
In terms of the (anti)-self-dual projections of the field, $A_{\pm} = \frac{1}{2} (1 \pm \star) A$ and $\dr \chi_{\pm} = \frac{1}{2} (1 \pm \star) \dr \chi$, the action becomes 
\begin{equation}
    \frac{\kappa}{2} \int_{\pd M} \! \big( 2 A_- \wedge A_- + 4 A_- \wedge \dr \chi_- - \dr \chi_+ \wedge \dr \chi_+ + \dr \chi_- \wedge \dr \chi_- \big) \ . 
\end{equation}
The self-dual component of $A$ has dropped entirely, and the anti-self-dual component is algebraically eliminated as $A_- = - \dr \chi_-$, giving the action 
\begin{equation}
    - \frac{\kappa}{2} \int_{\pd M} \! \dr \chi \wedge \dr \chi \ , 
\end{equation}
which evidently carries no degrees of freedom and vanishes when $\chi$ is global. Thus, we continue under the specification that $A$ is an odd-degree form (i.e.\ $n$ is odd), and $\Sigma \equiv \pd M$ is Lorentzian, such that $\star^2 (A \vert_{\pd M}) = A \vert_{\pd M}$. (Note that we could also include Euclidean signature when $n$ is odd by considering imaginary self duality, i.e.\ $\star A \vert_{\pd M} = i A \vert_{\pd M}$.)

In order to define a more general decomposition of the gauge field which will not break Lorentz invariance, we introduce a $1$-form $\omega \in \Omega^1 (\pd M)$ and its normalised dual vector $v = \omega^\sharp / \Vert \omega \Vert^2$ (such that $\iota_v \omega = 1$) and extend them to live on the whole manifold $M$. With this data, we can decompose the exterior derivative into 
\begin{equation}
    \dr = \dr^\perp + \dr^\para \ , \quad \dr^\para = \omega \wedge \CL_v \ , 
\end{equation}
such that $(\dr^\perp)^2 = 0$ and $(\dr^\para)^2 = 0$ when $\dr \omega = 0$, which we shall hence assume. The connection similarly decomposes as 
\begin{equation}
    A = A^\perp + \omega \wedge \iota_v A \ , \quad \iota_v A^\perp = 0 \ . 
\end{equation}
To make contact with the derivation of the FJ action from CS in section \ref{sec:CStoCB}, one could specify $\omega = \dr \tau$. Here, however, we keep $\omega$ arbitrary which will allow us to maintain Lorentz covariance in the resultant boundary theory.

Substituting this decomposition of the gauge field into our action \eqref{eq:Kasparov}, the bulk Chern-Simons term becomes 
\begin{equation}
    S_{\text{CS}}[A] = \kappa \int_{M} \big( 2 \omega \wedge \iota_v A \wedge \dr A^\perp + A^\perp \wedge \dr A^\perp \big) + \kappa \int_{\pd M} \omega \wedge \iota_v A \wedge A^\perp \ , 
\end{equation}
and $\iota_v A$ acts as a Lagrange multiplier enforcing the constraint $\omega \wedge \dr A^\perp = 0$. This has the general solution $A^\perp = \dr \phi - \omega \wedge C$ \cite[appendix C]{Bansal:2021bis} which is further fixed by the constraint $\iota_v A^\perp = 0$, implying $C = \iota_v \dr \phi$. We can therefore write the total gauge field as 
\begin{equation}
    A = \dr \phi - \omega \wedge \alpha \ , \quad \alpha \equiv \iota_v (\dr \phi - A) \ , 
\end{equation}
and the gauge invariant combination as $A^\chi = \dr b - \omega \wedge \alpha$ where $b = \phi + \chi$. Notice that this expression for $A^\chi$ can be identified with the combination we called $\mu$ in the introduction, and we will henceforth refer to it as such.

Substituting this back into the action \eqref{eq:Kasparov} gives 
\begin{equation} \label{eq:Kramnik}
    S_{\text{inv}} = - \frac{\kappa}{2} \int_{\pd M} \! \big( \mu \wedge \star \mu + 2 \omega \wedge \alpha \wedge \dr b \big) \ , 
\end{equation}
which may be expanded out as 
\begin{equation} \label{eq:Anand}
    S_{\text{inv}} = - \frac{\kappa}{2} \int_{\pd M} \! \big( \dr b \wedge \star \dr b + 2 \omega \wedge \alpha \wedge \CX + \omega \wedge \alpha \wedge \star (\omega \wedge \alpha) \big) \ , 
\end{equation}
where $\CX = \dr b - \star \dr b$. We recognise this as the sought-after Mkrtchyan action for chiral $p$-forms. We already know that this action comes equipped with a local symmetry, which is the uplift of the PST symmetry to the Mkrtchyan action, 
\begin{equation}
    \delta_{\epsilon} b = \epsilon \alpha \ , \quad \delta_{\epsilon} \omega = \dr \epsilon \ , \quad \delta_{\epsilon} \alpha = \epsilon \iota_v (1 - \star) \dr \alpha \ , 
\end{equation}
under which we have 
\begin{equation}
    \delta_{\epsilon} \mu = \epsilon (1 - \star) \iota_v (\omega \wedge \dr \alpha) = \epsilon (1 - \star) \iota_v \dr \mu \ . 
\end{equation}
Notice that this represents a \textit{zilch} symmetry: after imposing the constraint from the parallel component of the CS connection, we have a new symmetry proportional to $\dr \mu \equiv \dr A^\chi = \dr A$ which vanishes on-shell for both the CS and Mkrtchyan theories. Thus this symmetry can be understood as a trivial symmetry which arises once auxiliary fields are integrated out (in much the same way that SUSY closes only on-shell once auxiliaries are eliminated). To experts on these formalisms, this may not be particularly surprising as the PST symmetry is also known to be a \textit{zilch} symmetry in the same fashion \cite{Pasti:2012wv, Driezen:2016tnz}.

To go from the Mkrtchyan to the PST action we use the equation of motion for $\alpha$, 
\begin{equation}
    \omega \wedge \big( \CX + \star (\omega \wedge \alpha) \big) = 0 \ . 
\end{equation}
We can partially determine this solution as 
\begin{equation}
    \alpha = \iota_v \CX + \iota_v {\star} (\omega \wedge \rho) \, , 
\end{equation}
where $\rho$ is undetermined, but then, using the general identity $\iota_v \star \beta = \star (\beta \wedge v^\flat)$ and $v^\flat \equiv \omega / \Vert \omega \Vert^2$, we conclude that the second term does not contribute and $\alpha = \iota_v \CX$. Making use of a further identity $\star \iota_v \beta = (-1)^{p-1} v^\flat \wedge \star \beta$ for a general $p$-form $\beta \in \Omega^p (\pd M)$, we can entirely eliminate $\alpha$ from the action \eqref{eq:Anand} to recover the $p$-form version of the PST action, 
\begin{equation}
    S_{\text{inv}} = - \frac{\kappa}{2} \int_{\pd M} \! \big( \dr b \wedge \star \dr b - \Vert \omega \Vert^2 \iota_v \CX \wedge \star \iota_v \CX \big) \ . 
\end{equation}

\section{Comments on the double world sheet} 

Let us return momentarily to the two-dimensional Mkrtchyan action of eq. \eqref{eq:RuyLopez}, which we recast as 
\begin{equation} 
S_{\text{M}}[\phi, f, \alpha]   = \int_\Sigma \dr^2 \sigma \, \big( (\pd_+  \phi-  \alpha \pd_+f ) (\pd_-\phi    \alpha \pd_-f)   -  \alpha \pd_+f  \pd_- \phi +   \alpha \pd_-f \pd_+ \phi  \big) \ . 
\end{equation} 
There is another rather direct way to see the emergence of a Chern Simons description following the approach of \cite{Driezen:2016tnz}.  In \cite{Driezen:2016tnz} a similar first order formalism was obtained (in the context of non-linear sigma models) by invoking some partial gauge fixing of a two-dimensional gauge field $A_\pm = \alpha \partial_\pm f  $.   By reverse engineering we are led to consider an action,
\begin{equation} 
S'_{\text{M}}[\phi, A] = \int_\Sigma \dr^2 \sigma \, \big( (\pd_+  \phi- A_+) (\pd_-   \phi- A_-)   - A_+  \pd_- \phi +  A_-  \pd_+ \phi  \big) \ , 
\end{equation}
such that upon   fixing the aforementioned gauge, returns the Mkrtchyan Lagrangian.  In this way we have arrived at an action (upto a trivial integration by parts)  originally proposed by Witten \cite{Witten:1996hc} in a similar context.  This, however, is not gauge invariant under $\delta \phi = \epsilon $ and $\delta A_\pm = \partial_\pm \epsilon $ as 
\begin{equation} 
\delta S'_{\text{M}}[\phi,A] = \int_\Sigma \dr^2 \sigma \, \big(\epsilon ( \pd_- A_+ - \pd_+ A_-  \big) \ . 
\end{equation}
Thus,  now with opposite logic, to restore gauge invariance we add a Chern-Simons bulk action to arrive again at \eqref{eq:Kasparov}.  

Let us now consider the case where we have $2n$ bosonic fields, denoted by $\mathbb{X}$, that enter in a generalisation of the Mkrtchyan action:
\def\IX{\mathbb{X}}
\def\IA{\mathbb{A}}
\begin{eqnarray}
S &=&\frac 1 2\, \int d^2 \sigma \,\Big( (\partial _+f\IA-\partial _+\IX){\cal H }  (\partial _-f\IA-\partial _-\IX)+ 
\partial _-f\IA\eta\partial _+\IX -\partial _+f \IA\eta \partial _-\IX
\Big)\nonumber\,. 
\end{eqnarray}  
This is relevant to the doubled world sheet description of strings on a toroidal background where $\mathbb{X} = \{ x^i , \tilde{x}_i\}$ constitutes the coordinates of the target space together with their T-duals.  The   couplings are specified by a split signature pairing 
\begin{equation}
\eta=\left( \begin{array}{cc}0& 1 \\ 1 &0 \end{array}\right)\,,
\end{equation} 
and  a generalised metric encoding the target space metric and Kalb-Ramond data
\begin{equation}
{\cal H}=\left( \begin{array}{cc}g-bg^{-1}g&-bg^{-1}\\g^{-1}b&g^{-1} \end{array}\right) \, .
\end{equation}  
Note that the generalised metric is an element of $O(n,n)$ (the transformations that preserve $\eta$) and so defines an almost product structure  ${\mathcal E} = {\cal H}\eta^{-1}$ that obeys ${\mathcal E}^2 = 1$.     Elimination of the scalar fields $\mathbb{A} $ from the action yields a PST formulation of the doubled string, as in \cite{Driezen:2016tnz}, from which   twisted self-duality  $d\mathbb{X} = \star {\mathcal E} d \mathbb{X}$ follows as the equation of motion.  From this twisted self-duality constraint one can eliminate half the variables, the $\tilde{x}_i$ say, to yield second order equations for the other half, the $x^i$ which reproduce those of the standard string world sheet.

We now follow the same strategy as above and propose to undo a gauge fixing $\IA_\pm = \IA  \partial_\pm f$ by considering 
\begin{equation}
S[\mathbb{A}] =\frac 1 2\, \int d^2 \sigma \,\Big( (\IA_+-\partial _+\IX){\cal H }  (\IA_--\partial _-\IX)+ \partial _+\IX \eta\IA_--
\partial _-\IX \eta\IA_+\Big)
\,.\label{mk22}
\end{equation}
This action is not gauge invariant under the full $U(1)^{2n}$ symmetry $\delta \IA= d\epsilon$, $\delta \IX = \epsilon$.  It is however invariant when we consider the gauge fields to take values in an $U(1)^n$ sub-algebra that is isotropic with respect to $\eta$, i.e. if we gauge half the symmetries with $\eta_{IJ} \mathbb{A}^I \mathbb{A}^J =\eta_{IJ} \epsilon^I \epsilon^J =\eta_{IJ} \epsilon^I \mathbb{A}^J =   0 $.  This idea of gauging an isotropic sub-algebra was invoked by Hull in his approach to the doubled string \cite{Hull:2006va} and developed in  \cite{Lee:2013hma}.   If instead we wish to restore a full $U(1)^{2n}$ invariance we add a Chern-Simons term to arrive at the form  
\begin{equation} \label{eq:CSEmodel}
    S_{\text{inv}}[\mathbb{A}, \mathbb{\IX }] =  S_{\text{CS}}[\mathbb{A}] + \int_{\pd M} \! \eta \big( \mathbb{A} \wedge \dr \IX  - \tfrac{1}{2} (\mathbb{A} -\dr \IX ) \wedge
   \star  {\cal E }  (\mathbb{A} -\dr \IX ) \big)  \ . 
\end{equation} 
  where $U(1)^{2n}$ algebra indices are contracted with the product $\eta$.    This discussion can be expanded to the case where T-duality acts in an internal space that is a fibration over some base manifold.  To do so one simply allows ${\mathcal H} = {\mathcal H} (y) $  to depend on the coordinates $y^a$ of the base manifold and couple to a background gauge field $\mathbb{B} = \mathbb{B}_a dy^a$, an $O(n,n)$ vector that contains {\em off-diagonal} metric and two-form data (see \cite{Driezen:2016tnz} eq. 2.3),  by making a minimal coupling substitution  $\dr \mathbb{X} \rightarrow \nabla \mathbb{X} = \dr \mathbb{X} - \mathbb{B}$.

\section{Non-Abelian Chern-Simons}

Having seen that our formalism reduces to others found in the literature, we will now leverage the simplicity of our approach to provide a novel generalisation. By starting with non-Abelian Chern-Simons theory in $3$-dimensions, we will derive a polynomial action for non-Abelian chiral bosons. This will be the non-Abelian generalisation of the Mkrtchyan action and integrating out an auxiliary field will yield the PST action for non-Abelian chiral bosons.

Let $G$ be a Lie group whose algebra $\fg$ is equipped with an ad-invariant inner product $\biprod{\bullet}{\bullet}$, and consider the algebra-valued $1$-forms $A \in \Omega^1 (M, \fg)$. The action to consider is 
\begin{equation} \label{eq:Larsen}
    S^\prime_{\text{CS}}[A] = \int_{M} \big( \biprodf{A}{\dr A} + \frac{1}{3} \triprodf{A}{A}{A} \big) - \frac{1}{2} \int_{\pd M} \! \biprodf{A}{\star A} \ , 
\end{equation}
which is the non-Abelian CS action plus a boundary term which renders the boundary condition $A \vert_{\pd M} = \star (A \vert_{\pd M})$ Neumann.

As in the Abelian case, the presence of a boundary spoils the gauge invariance under $A \to A^g \equiv g^{-1} A g + g^{-1} \dr g$, but we may restore the invariance of the action by coupling the bulk CS theory to a boundary edge mode. Let $h \in C^\infty(\pd M, G)$ be a boundary field which transforms as $h \to g^{-1} h$ such that $A^h \equiv h^{-1} A h + h^{-1} \dr h$ is invariant. Then, as before, we simply replace $A \to A^h$ in our action to find the gauge invariant theory, 
\begin{equation} \label{eq:Magnus}
    S_{\text{inv}}[A, h] \equiv S^\prime_{\text{CS}}[A^h] = S_{\text{CS}}[A] + S_{\text{WZ}}[h] + \int_{\pd M} \! \big( \biprodf{A}{\dr h h^{-1}} - \tfrac{1}{2} \biprodf{A^h}{\star A^h} \big) \ , 
\end{equation}
where the Wess-Zumino [WZ] term is defined by 
\begin{equation}
    S_{\text{WZ}}[h] = - \frac{1}{6} \int_{M} \triprodf{h^{-1} \dr h}{h^{-1} \dr h}{h^{-1} \dr h} \ . 
\end{equation}

As the action \eqref{eq:Magnus} is now gauge invariant under the bulk transformation, one might wonder how the quantisation of the CS level arises since the conventional argument about large gauge transformations is rendered moot. The answer is simply that we are required to extend the edge mode $h$ into the bulk to define the WZ term, and, in the standard fashion, demanding that the path integral is insensitive to the choice of such an extension invokes the desired quantisation condition.

Following the same recipe, we introduce a $1$-form $\omega \in \Omega^1 (\pd M)$ and its normalised dual vector $v = \omega^\sharp / \Vert \omega \Vert^2$ (such that $\iota_v \omega = 1$), which we extend to live on the whole manifold $M$. With this data, we can decompose the exterior derivative into 
\begin{equation}
    \dr = \dr^\perp + \dr^\para \ , \quad \dr^\para = \omega \wedge \CL_v \ , 
\end{equation}
such that $(\dr^\perp)^2 = 0$ and $(\dr^\para)^2 = 0$ when $\dr \omega = 0$, which we shall hence assume. The connection similarly decomposes as 
\begin{equation}
    A = A^\perp + \iota_v A \, \omega \ , \quad \iota_v A^\perp = 0 \ , 
\end{equation}
where $\iota_v A \in C^\infty(M, \fg)$ is now valued in the algebra.

Under this decomposition, the bulk CS term becomes 
\begin{equation}
    S_{\text{CS}}[A] = \int_{M} \big( 2 \omega \wedge \biprod{\iota_v A}{F^\perp} + \biprodf{A^\perp}{\dr A^\perp} \big) + \int_{\pd M} \! \omega \wedge \biprod{\iota_v A}{A^\perp} \ , 
\end{equation}
where $F^\perp = \dr A^\perp + A^\perp \wedge A^\perp$, and we see that $\iota_v A$ is again acting as a Lagrange multiplier fixing 
\begin{equation}
    \omega \wedge F^\perp = 0 \ . 
\end{equation}
In order to write the most general solution to this equation we will need to slightly generalise the argument presented in \cite[appendix C]{Bansal:2021bis} to non-Abelian fields. Let us assume that the closed $1$-form $\omega$ is sufficiently \textit{nice}, meaning $\omega \sim \dr f$ where $f$ is a good global coordinate and we can foliate our $3$-manifold by slices of constant $f$. On each slice of constant $f$, the constraint above reduces to a non-Abelian flatness condition $F^\perp = 0$, which we can solve with $A^\perp = g^{-1} \dr g$ using the non-Abelian Poincar{\' e} lemma. We can now glue these solutions back together to form the solution on the whole manifold, and the only piece we might have missed is a component parallel to $\omega$. Thus, the most general solution is 
\begin{equation}
    A^\perp = g^{-1} \dr g - C \, \omega \ . 
\end{equation}
We can fix this solution further by imposing the constraint $\iota_v A^\perp = 0$, implying $C = \iota_v (g^{-1} \dr g)$. We can therefore write the total gauge field as 
\begin{equation}
    A = g^{-1} \dr g - h \alpha h^{-1} \, \omega \ , \quad \alpha \equiv h^{-1} \iota_v (g^{-1} \dr g - A) h \ , 
\end{equation}
and the gauge invariant combination is written as $A^h = m^{-1} \dr m - \alpha \, \omega$ where $m = gh$. To match notation with the previous section, we will again relabel this combination as $\mu \equiv A^h$.

Substituting this solution back into the action \eqref{eq:Magnus}, we find that it may be succinctly expressed as 
\begin{equation} \label{eq:Niemann}
    S_{\text{inv}} = S_{\text{WZ}}[m] - \frac{1}{2} \int_{\pd M} \! \big( \biprodf{\mu}{\star \mu} + 2 \omega \wedge \biprod{\alpha}{m^{-1} \dr m} \big) \ . 
\end{equation}
It is actually easiest to compute this action by substituting the expression for $\mu \equiv A^h$ directly into eq.\ \eqref{eq:Larsen}. Defining $\CX = m^{-1} \dr m - \star (m^{-1} \dr m)$, this may be expanded out as 
\begin{equation} \label{eq:Giri}
    S_{\text{inv}} = S_{\text{WZW}}[m] - \frac{1}{2} \int_{\pd M} \! \big( 2 \omega \wedge \biprod{\alpha}{\CX} + \biprod{\alpha}{\alpha} \, \omega \wedge \star \omega \big) \ , 
\end{equation}
to provide the non-Abelian generalisation of the Mkrtchyan action \eqref{eq:Anand}.

This action is invariant under the local transformation 
\begin{equation} \label{eq:Polgar}
    \delta_{\epsilon} m = \epsilon m \alpha \ , \quad \delta_{\epsilon} \omega = \dr \epsilon \ , \quad \delta_{\epsilon} \alpha = \epsilon \iota_v (1 - \star) \nabla \alpha \ , 
\end{equation}
where $\nabla \bullet = \dr \bullet + [m^{-1} \dr m, \bullet]$. To demonstrate this invariance, it is helpful to use the action \eqref{eq:Niemann} and note that $\delta (\alpha \, \omega) = \nabla (\epsilon \alpha) - \delta \mu$ and $\omega \wedge \delta \mu = \epsilon \, \omega \wedge \nabla \alpha$. Also, the covariant derivative $\nabla$ is nilpotent, i.e.\  $\nabla^2 = 0$, and satisfies 
\begin{equation}
    \int_{\pd M} \! \biprodf{A_1}{\nabla A_2} = (-1)^{\deg A_1 + 1} \int_{\pd M} \! \biprodf{\nabla A_1}{A_2} 
\end{equation}
for any pair of $\fg$-valued forms $A_1$, $A_2$. The calculation showing that the action is invariant under these transformations is given in more detail in appendix \ref{app:NAinv}.

As in the Abelian case, this symmetry is a \textit{zilch} symmetry arising from the elimination of auxiliary fields. This can be most easily seen by computing 
\begin{equation}
    \delta_{\epsilon} \mu = \epsilon (1 - \star) \iota_v (\omega \wedge \nabla \alpha) = \epsilon (1 - \star) \iota_v (\dr \mu + \mu \wedge \mu) \ , 
\end{equation}
and the right hand side is proportional to the field strength of $\mu \equiv A^h$, related to the regular CS field strength by $F[\mu] \equiv F[A^h] = h^{-1} F[A] h$, which vanishes on-shell.

The elimination of $\alpha$ by its equation of motion proceeds as in the Abelian case, \textit{mutatis mutandis}, to yield the non-abelian PST action, 
\begin{equation}
    S_{\text{inv}} = S_{\text{WZW}}[m] + \frac{1}{2} \int_{\pd M} \! \star \Vert \omega \Vert^2 \biprod{\iota_v \CX}{\iota_v \CX} \ , 
\end{equation}
and setting $\omega = \dr \tau$ gives the FJ-type action for non-Abelian chiral bosons \cite{Sonnenschein:1988ug}, 
\begin{equation} \label{eq:Nimzo}
    S_{\text{inv}} = - \int_{\pd M} \! \dr^2 \sigma \, \biprod{m^{-1} \pd_\sigma m}{m^{-1} \pd_- m} + S_{\text{WZ}}[m] \ . 
\end{equation}

\subsection{Self-dual gauge transformations as   affine transformations}
 \label{sec:affine}

In the above, we saw that our coupled bulk-boundary action \eqref{eq:Magnus} was invariant under the  conventional transformation for the gauge field combined with a left action on the edge mode, 
\begin{equation}
    A \to A^g \equiv g^{-1} A g + g^{-1} \dr g \ , \quad h \to g^{-1} h \ , \quad A^h \to A^h \ . 
\end{equation}
We can, however, consider a further set of local transformations which follow from the \textit{right} action on $h$ leaving $A$ invariant, 
\begin{equation}
    A \to A \ , \quad h \to h g \ , \quad A^h \to A^{hg} \equiv g^{-1} A^h g + g^{-1} \dr g \ . 
\end{equation}
Under this right action, the action \eqref{eq:Magnus} transforms as 
\begin{equation}\label{alsoPolgar}
\begin{split}
    S_{\text{inv}}[A, h] \to  S_{\text{inv}}[A, h]+ S_{\text{WZ}}[g] & + \int_{\pd M} \! \biprodf{A^h}{(1 - \star)\dr g g^{-1}} \\ 
    & \hspace{5mm} - \frac{1}{2} \int_{\pd M} \biprodf{\dr g g^{-1}}{\star (\dr g g^{-1})} \ . 
\end{split}
\end{equation}
For self-dual ``gauge'' transformations, i.e.\ those which obey $\dr g g^{-1} = \star (\dr g g^{-1})$, the two boundary terms are zero. Notice that this condition may also be written as $\pd_- g g^{-1} = 0$. In order to kill the WZ term, we assume that the bulk extension of $g$ can be suitably chosen so as to also obey this constraint. Pushed through to eq.\ \eqref{eq:Nimzo} these transformations correspond to an affine right action, 
\begin{equation}
    m \to m g(\sigma^+) \ , 
\end{equation}
and gives rise to a chiral current algebra \cite[page 18]{Sonnenschein:1988ug}.

To see that these self-dual boundary transformations are in fact affine transformations and not gauge symmetries, we should compute their Noether charges. First, working directly at the level of the chiral WZW model, eq.\ \eqref{eq:Nimzo}, we compute the Noether charge for the infinitesimal affine right action to be 
\begin{equation}
    Q = 2 \int \dr \sigma \, \biprod{\varepsilon}{m^{-1} \pd_\sigma m} \ . 
\end{equation}
The fact that this charge is non-vanishing implies that these transformations are not gauge symmetries of the theory. Furthermore, we can compute the same Noether charge in the Chern-Simons theory, 
\begin{equation}
    Q = \int \dr \sigma \, \biprod{\varepsilon}{(A_\tau + A_\sigma)} \ , 
\end{equation}
which comes from the conserved current $J = A + \star A$. If we impose the boundary condition $A_\tau = A_\sigma$ and substitute in the solution to the equation of motion, we see that these Noether charges agree on-shell. 

Finally, we observe that this same calculation generalises immediately to the case of an abelian chiral $p$-form field (with self-dual $(p+1)$-form field strength in $2(p+1)$ dimensions) and yields a global ``affine'' symmetry of the chiral $p$-form gauge theory, with (nonvanishing) $p+1$-form Noether current $  J=A+\star A $ as before.

\section{Twisted Self-Duality and PLTD}

One further generalisation which we wish to consider is the possibility for a twisted self-duality relation of the form $A \vert_{\pd M} = \CE {\star} (A \vert_{\pd M})$. This boundary condition has been considered recently in the context of PLTD \cite{Severa:2016prq} and 4d CS theory \cite{Lacroix:2020flf}. Given our choice of boundary $\Sigma = \pd M$ such that\footnote{Note that we could also include Euclidean signature by considering imaginary self duality, i.e.\ $\star (A \vert_{\pd M}) = i A \vert_{\pd M}$ and $\star^2 (A \vert_{\pd M}) = - A \vert_{\pd M}$.} $\star^2 (A \vert_{\pd M}) = A \vert_{\pd M}$, we have the self-consistency condition $\CE^2 = 1$, and therefore we should think of $\CE$ as an involution on the algebra. For many purposes we might take $\CE$ to be constant (though this is not essential and indeed there are natural examples in which $\CE$ could be a function of other fields in a larger system) and the above relation can be understood to hold pointwise.

Generalising our formalism to include twisted self-duality is actually rather straight forward. Instead of adding the boundary term $A \wedge \star A$ to the CS action, we add the term $A \wedge \CE {\star} A$, and the rest of the analysis follows through as before, so long as we impose the condition $\biprod{\bullet}{\CE \bullet} = \biprod{\CE \bullet}{\bullet}$. This can be thought of as a compatibility condition between the involution $\CE$ and the inner product $\biprod{\bullet}{\bullet}$, or alternatively as the requirement that $(\bullet \, , \bullet) = \int_\Sigma \biprod{\bullet}{\CE {\star} \bullet}$ is a symmetric inner product on algebra-valued 1-forms.

So, taking this compatibility condition as given, and starting from the action 
\begin{equation}
    S_{\text{inv}}[A, h] = S_{\text{CS}}[A] + S_{\text{WZ}}[h] + \int_{\pd M} \! \big( \biprodf{A}{\dr h h^{-1}} - \tfrac{1}{2} \biprodf{A^h}{\CE {\star} A^h} \big) \ , 
\end{equation}
our derivation culminates in 
\begin{equation}
\begin{split}
    S_{\text{inv}} = S_{\text{WZ}}[m] - \frac{1}{2} \int_{\pd M} \! \biprodf{m^{-1} \dr m}{\CE {\star} (m^{-1} \dr m)} & - \int_{\pd M} \! \omega \wedge \biprod{\alpha}{\CX} \\ 
    & - \frac{1}{2} \int_{\pd M} \! \biprod{\alpha}{\CE \alpha} \, \omega \wedge \star \omega \ , 
\end{split}
\end{equation}
where $\CX = m^{-1} \dr m - \CE {\star} (m^{-1} \dr m)$. This is the non-abelian Mkrtchyan action, eq.\ \eqref{eq:Giri}, but now with twisted self-duality.

The elimination of $\alpha$ by its equation of motion proceeds directly to give the non-abelian twisted PST action, 
\begin{equation} \label{eq:Hikaru}
    S_{\text{inv}} = S_{\text{WZ}}[m] - \frac{1}{2} \int_{\pd M} \! \biprodf{m^{-1} \dr m}{\CE {\star} (m^{-1} \dr m)} + \frac{1}{2} \int_{\pd M} \! \star \Vert \omega \Vert^2 \biprod{\iota_v \CX}{\iota_v \CX} \ . 
\end{equation}
Introducing projectors $\CP_{\pm} = \frac{1}{2} (1 \pm \CE)$ we can write $\iota_v \CX$ in lightcone coordinates as 
\begin{equation}
    \iota_v \CX = \frac{1}{\Vert \omega \Vert^2} \big( \omega_- \CP_- (m^{-1} \pd_+ m) + \omega_+ \CP_+ (m^{-1} \pd_- m) \big) \ , 
\end{equation}
and substituting this into eq.\ \eqref{eq:Hikaru} gives 
\begin{equation}
\begin{split}
    S = S_{\text{WZ}}[m] & - \frac{1}{2} \int_{\pd M} \! \dr^2 \sigma \, \biprod{m^{-1} \pd_+ m}{\CE (m^{-1} \pd_- m)} \\ 
    & + \frac{1}{2} \int_{\pd M} \! \dr^2 \sigma \, \frac{\omega_+}{\omega_-} \biprod{m^{-1} \pd_- m}{\CP_+ (m^{-1} \pd_- m)} \\
    & - \frac{1}{2} \int_{\pd M} \! \dr^2 \sigma \, \frac{\omega_-}{\omega_+} \biprod{m^{-1} \pd_+ m}{\CP_- (m^{-1} \pd_+ m)} \ . 
\end{split}
\end{equation}
This is now easily compared with \cite[eq.\ 3.23]{Driezen:2016tnz}.

Finally, fixing $\omega = \dr \tau$ returns the FJ form with twisted self-duality, 
\begin{equation} \label{eq:Ding}
    S = S_{\text{WZ}}[m] - \int_{\pd M} \! \dr^2 \sigma \, \big( \biprod{m^{-1} \pd_\sigma m}{m^{-1} \pd_\tau m} - \biprod{m^{-1} \pd_\sigma m}{\CE (m^{-1} \pd_\sigma m)} \big) \ . 
\end{equation}
In the special case where the algebra is a Drinfeld double $\fd = \fg \bowtie \tilde{\fg}$, and the inner product is the natural pairing $\eta = \biprod{\bullet}{\bullet}$, the compatibility condition $\biprod{\bullet}{\CE \bullet} = \biprod{\CE \bullet}{\bullet}$ is precisely that which appears in the context of Poisson-Lie T-duality. In that context, the FJ form \eqref{eq:Ding} is often denoted the $\CE$-model \cite{Klimcik:1995ux, Klimcik:1996nq, Klimcik:1995dy}.

One can now consider the transformations which gave rise to affine symmetries in the untwisted case. In general, one expects a non-trivial choice of $\CE$ to reduce the symmetries of the theory. Indeed, in order for the action to be invariant under $m \to m g$, the gauge parameter $g$ is required to obey 
\begin{equation}\label{eq:Capablanca}
    \CE = \mathrm{Ad}_g \circ \CE \circ \mathrm{Ad}_{g^{-1}} \quad \text{and} \quad \dr g g^{-1} = \CE {\star} (\dr g g^{-1}) \ . 
\end{equation}
Given these constraints, we can compute the Noether charge in the Chern-Simons model, 
\begin{equation}
    Q = \int \dr \sigma \, \biprod{\varepsilon}{(A_\tau + \CE A_\sigma)} \ , 
\end{equation}
which comes from the conserved current $J = A + \CE {\star} A$. As before, the fact that this charge is non-vanishing tells us that this is not a gauge symmetry for the theory. Instead, we should interpret it as a twisted affine symmetry. 

As an example of the above it is informative to consider the case where $\fd = \fg  \oplus \fg $ for which isotropic spaces of $\langle \bullet, \bullet \rangle$ are the diagonal and anti-diagonal subsets (of which only the diagonal is a Lie sub-algebra).   This case is relevant to the construction of the WZW model, and its integrable $\lambda$-deformation \cite{Klimcik:2015gba}.  We define $t_a$ as generators of $\fg$ with $f_{ab}{}^{c}$ the structure constants and $\kappa_{ab}$ the Cartan-Killing metric.  A basis of $\fd$ is formed by $T_a =\{ t_a , - t_a\} $ and $\tilde{T}^a =\kappa^{ab}\{  t_a ,  t_a \}$ spanning the two isotropics and we specify $\mathcal{E}$ by setting $\mathcal{E}(T_a) = \mu  \, \kappa_{ab}  \tilde{T}^b$.    In this case the algebraic part of condition eq.~\eqref{eq:Capablanca}, working infinitesimally in  the affine symmetry parameter $g = \exp(\xi) = \exp( \xi^a T_a + \tilde{\xi}_a \tilde{T}^a)$, yields one non-trivial constraint
\begin{equation}
    0= (\mu^{-1} - \mu) f_{ab}{}^c \xi^a 
\end{equation}
which is also trivially solved, with no further condition on $\xi$, when $\mu = 1$ (corresponding to the un-deformed WZW model). The differential condition in eq.~\eqref{eq:Capablanca} then implies $\dr \xi^a = \star \kappa^{ab} \dr \tilde{\xi}_b$, or equivalently that $\xi = \{\xi_L(\sigma^+)  , \xi_R(\sigma^-) \} $.  The ${\cal E}$-model descends to a WZW model, and its $\lambda$-deformation parameterised by $\mu$, upon reduction to the coset $G_{\textrm{diag}}  
 \diagdown \exp(\fd) $ defined by the equivalence relation $\{g_1 , g_2 \} \sim \{ h g_1 , h g_2 \} $.   We choose a coset representative $m= \{1, h\}$ for a group element $h$ which will be the field of the WZW model. We see that the affine symmetry acts as 
\begin{equation}
    m \mapsto \{ h_L(\sigma^+) , h h_R(\sigma^-) \} \sim \{ 1  , h^{-1}_L(\sigma^+) h h_R(\sigma^-) \} \ . 
\end{equation}
Thus the twisted chiral affine transformation on the double algebra descends to the anticipated $G_L \times G_R $ affine symmetry of the WZW model.

\section*{Acknowledgements}

ASA, OH, AS, and DCT are supported by the FWO-Vlaanderen through the project G006119N, as well
as by the Vrije Universiteit Brussel through the Strategic Research Program ``High-Energy Physics''. ASA is also supported by an FWO Senior Postdoctoral Fellowship (number 1265122N).    DCT is supported by The Royal Society through a University Research Fellowship
Generalised Dualities in String Theory and Holography URF 150185 and in part by STFC
grant ST/P00055X/1. For the purpose of open access, the authors have applied a Creative Commons Attribution (CC BY) licence to any Author Accepted Manuscript version arising.

\appendix

\section{Invariance of the non-Abelian Mkrtchyan action} \label{app:NAinv}

In this appendix, we will demonstrate that the action \eqref{eq:Niemann}, 
\begin{equation} 
    S_{\text{inv}} = S_{\text{WZ}}[m] - \frac{1}{2} \int_{\pd M} \! \big( \biprodf{\mu}{\star \mu} + 2 \omega \wedge \biprod{\alpha}{m^{-1} \dr m} \big) \ , 
\end{equation}
is invariant under the local transformation \eqref{eq:Polgar} 
\begin{equation}
    \delta_{\epsilon} m = \epsilon m \alpha \ , \quad \delta_{\epsilon} \omega = \dr \epsilon \ , \quad \delta_{\epsilon} \alpha = \epsilon \iota_v (1 - \star) \nabla \alpha \ , 
\end{equation}
where $\nabla \bullet = \dr \bullet + [m^{-1} \dr m, \bullet]$ is the covariant derivative with respect to the flat connection $m^{-1} \dr m$. The covariant derivative $\nabla$ is nilpotent $\nabla^2 = 0$ and satisfies 
\begin{equation}
    \int_{\pd M} \! \biprodf{A_1}{\nabla A_2} = (-1)^{\deg A_1 + 1} \int_{\pd M} \! \biprodf{\nabla A_1}{A_2} 
\end{equation}
for any pair of $\fg$-valued forms $A_1$, $A_2$. In terms of this operator we have a general formula for the variation of $m^{-1}\dr m$:
\be
\delta_{\epsilon} (m^{-1}\dr m)=\nabla(m^{-1}\delta m)\,.
\ee

From \eqref{eq:Polgar} the last formula gives $\delta_{\epsilon} (m^{-1}\dr m)=\nabla(\epsilon \alpha)$. We have
\be
\delta_{\epsilon} S_{\rm WZ}[m]=-\int_{\pd M}\langle \epsilon \alpha,\nabla(m^{-1}\dr m)\rangle
\ee
and the variation of the remainder --- that must cancel against this --- reads
\be
-\int_{\pd M} \langle\delta_{\epsilon} \mu, \star \mu\rangle + \langle \delta_{\epsilon}(\alpha\omega) , m^{-1}\dr m \rangle + \langle\alpha \omega, \nabla(\epsilon \alpha)\rangle
\ee
The expression $\delta_{\epsilon}(\alpha\omega)$ is terrible and we will cancel it. We use $\star \delta_{\epsilon} \mu=-\delta_{\epsilon} \mu$ to simplify the first term inside the integral to $-\langle\mu,\delta_{\epsilon} \mu\rangle$. Then the first and last terms together yield
\begin{align}
-\mu \delta_{\epsilon} \mu + \alpha \omega \nabla(\epsilon \alpha)&=(m^{-1}\dr m) \delta_{\epsilon} \mu -\alpha \omega \delta_{\epsilon} \mu+\alpha \omega \nabla(\epsilon\alpha)\,,\\
&=(m^{-1}\dr m) \delta_{\epsilon} \mu-\epsilon \alpha \omega \nabla \alpha +\alpha \omega\nabla(\epsilon\alpha)\,;
\end{align}
the last two terms vanish by integration by parts (using $\nabla\omega=\dr \omega=0$). Therefore
\begin{equation}
\begin{split}
-\mu \delta_{\epsilon} \mu + \alpha \omega \nabla(\epsilon \alpha) & = (m^{-1}\dr m) \delta_{\epsilon} \mu \\
& =- (m^{-1}\dr m)\nabla(\epsilon \alpha) + (m^{-1}\dr m) \delta_{\epsilon}(\alpha\omega)\,.
\end{split}
\end{equation}
This last term cancels $\delta_{\epsilon}(\alpha\omega)m^{-1}\dr m$ inside the remainder; the latter reduces to
\be
+\int_{\pd M}\langle m^{-1}\dr m),\nabla(\epsilon \alpha)\rangle=-\int_{\pd M}\langle \nabla(\epsilon \alpha),m^{-1}\dr m\rangle\,,
\ee
which indeed cancels the variation of $S_{\rm WZ}[m]$.

\printbibliography

\end{document}